\documentstyle[prl,twocolumn,aps,epsf]{revtex}
\begin{document}
\draft

\tolerance 50000

\twocolumn[\hsize\textwidth\columnwidth\hsize\csname@twocolumnfalse\endcsname

\title{A Density Matrix Renormalization Group Study of \\
Ultrasmall Superconducting Grains
}

\author{J. Dukelsky$^{1}$ and  G. Sierra$^{2}$}

\address{
$^{1}$Instituto de Estructura de la Materia, C.S.I.C.,Madrid, Spain.
\\
$^{2}$Instituto de Matem{\'a}ticas y F{\'\i}sica Fundamental, C.S.I.C.,
Madrid, Spain.
}

\maketitle

\begin{abstract}
\begin{center}
\parbox{14cm}{
We apply the DMRG method to the BCS pairing Hamiltonian
which describes ultrasmall superconducting grains.
Our version of the DMRG uses the particle (hole) states
around the Fermi level
as the system block (environment).
We observe a smooth logarithmic-like  crossover
between the few electron regime and the BCS-bulk regime.
}

\end{center}
\end{abstract}

\pacs{
\hspace{2.5cm}
}

%\vskip2pc] \narrowtext

\vskip1pc] \narrowtext

The Density Matrix Renormalization Group (DMRG) \cite{DMRG} has been applied
with great success to a large variety of systems in Condensed Matter and
Statistical Mechanics (see \cite{Dresden} for a review). Most of these
applications use the real space formulation although the DMRG can also be
formulated in momentum space \cite{Xiang}. The starting point of the DMRG
method is the breaking of the system into two pieces, the block and the
environment, which are described by a finite number of states out of which
one can reconstruct the ground state and the excited states of the whole
system. A correct choice of the block and the environment is therefore
crucial for the DMRG to work. For example open chains are divided into left
and right handed pieces linked by a couple of sites. The local interaction
between the left and right pieces of the chains explains the adequacy of the
DMRG in 1d problems. However for more complicated systems, or in dimensions
higher than one, there are no general rules dictating the correct DMRG
breaking except the nature of the physical problem under study.

In this letter we shall present a DMRG analysis of the ground state (GS) of
the BCS Hamiltonian which is used to describe the superconducting properties
of ultrasmall Al grains, discovered by Ralph, Black and Tinkham \cite{RBT}.
We shall show that the DMRG gives an accurate approximation to the exact GS
if the block is taken to be the set of particles while the environment is
taken to be the set of holes. This choice does not satisfy the local
interaction rule, which is so effective in real space, for particles and
holes are all coupled by the BCS Hamiltonian. Nevertheless the projection of
the GS into the particle and hole subspaces via density matrices is strongly
peaked on a small number of states, which explains the applicability of the
DMRG to this problem.

The BCS pairing Hamiltonian used for small metallic grains is given by \cite
{janko,vD,Braun1,BvD,ML,MFF,Braun2}

\begin{equation}
H = \sum_{j=1,\sigma= \pm}^{\Omega} (\epsilon_j -\mu) c^\dagger_{j,\sigma}
\; c_{j,\sigma} - \lambda d \sum_{i,j=1}^{\Omega} \; c^\dagger_{i,+}
c^\dagger_{i,-} c_{j,-} c_{j,+}  \label{1}
\end{equation}

\noindent where $i,j=1,2,\dots ,\Omega $ label single particle energy levels
whose energies are given for simplicity by $\epsilon _{j}=jd$, and $d$ is
the average level spacing which is inversely proportional to the size of the
grain. $c_{j,\sigma }$ are electron destruction operators of time reserved
states $\sigma =\pm $. Finally $\mu $ is the chemical potential and $\lambda 
$ is the BCS coupling constant, whose appropriate value for the Al grains is
0.224 \cite{BvD}. Given $N$ electrons they can form $P$ Cooper pairs and $b$
unpaired states such that $N=2P+b$. The Hamiltonian (\ref{1}) decouples the
unpaired electrons and hence $b$ is a conserved quantity. We shall
investigate in this letter the GS of even grains ($b=0$) and odd grains ($b=1
$) at half filling ( $N=\Omega $) using a new version of the DMRG and
compare our results with the exact Lanczos diagonalization for $N=24$ and
projected BCS (PBCS) for larger $N$.

The Hamiltonian (\ref{1}) has two regimes depending on the ratio $d/\Delta=
2 \; {\rm sinh}(1/\lambda)/N $, between the level spacing $d$ and the bulk
superconducting gap $\Delta$ \cite{janko,vD,Braun1,BvD,ML,MFF,Braun2}. In
the weak coupling region ($d/\Delta >>1$), which corresponds to small grains
or small coupling constants, the system is in a regime with strong pairing
fluctuations above the Fermi sea which lead to logarithmic renormalizations 
\cite{ML}. In the strong coupling regime ($d/\Delta <<1$), which corresponds
to large grains or strong coupling constants, the bulk-BCS wave function
describes correctly the GS properties. In mean field theory the crossover
between the weak and strong coupling regimes occurs at a critical value of
the level spacing $d^C_b$ which is parity dependent. For even grains one has 
$d^C_0 /\Delta = 3.56$ while for odd grains $d^C_1/\Delta = 0.89$ \cite{vD}.

It is illustrative to consider the case where $\epsilon_j=0, \forall j $, so
that the exact GS for $b=0$ and $N= \Omega$ is given by the PBCS state

\begin{eqnarray}
&|N=\Omega,b=0, \epsilon_j=0 \rangle = \frac{1}{\sqrt{Z_{\Omega/2,\Omega}}}
\left( P^\dagger \right)^{\Omega/2} |0 \rangle &  \label{2} \\
&P^\dagger = \sum_{i=1}^\Omega c^\dagger_{i,+} c^\dagger_{i,-} &  \nonumber
\end{eqnarray}

\noindent where $|0\rangle $ is the vacuum of the electron operators and $%
Z_{M,N}=(M!)^{2}\;C_{N,M}\;(C_{N,M}=\left( 
\begin{array}{cc}
N &  \\ 
M & 
\end{array}
\right) )$ is the norm of the state $(P^{\dagger })^{M}|0\rangle $. Notice
that (\ref{2}) is the projection of the BCS state ${\rm exp}(P^{\dagger
})|0\rangle $ into the subspace with $\Omega $ electrons. On the other hand
for $\lambda =0$ and $\epsilon_j = j d$ the GS at half filling is a Fermi
sea with all the levels between $i=1$ and $i=N/2$ occupied. This state can
be written as

\begin{eqnarray}
&|FS\rangle =\prod_{i=1}^{\Omega /2}c_{i,+}^{\dagger }c_{i,-}^{\dagger
}|0\rangle =\frac{1}{\sqrt{Z_{\Omega /2,\Omega /2}}}\;\left( B^{\dagger
}\right) ^{\Omega /2}|0\rangle &  \label{3} \\
&B^{\dagger }=\sum_{i=1}^{\Omega /2}c_{i,+}^{\dagger }c_{i,-}^{\dagger }& 
\nonumber
\end{eqnarray}

Comparing (\ref{2}) and (\ref{3}) it is clear that the PBCS state (\ref{2})
can be derived from the state (\ref{3}) acting with pairs of particle-hole
(p-h) creation operators. With this aim we split the operator $P^{\dagger }$
as $A^{\dagger }+B^{\dagger }$, where $A^{\dagger }=\sum_{i=\Omega
/2+1}^{\Omega }c_{i,+}^{\dagger }c_{i,-}^{\dagger }$. After some algebra one
finds \cite{Schuck},

\begin{eqnarray}
&|N=\Omega,b=0, \epsilon_j=0 \rangle = \sum_{\ell=0}^{\Omega/2} \psi_{\ell}
\; |\ell \rangle_p \otimes |\ell \rangle_h &  \nonumber \\
& \psi_\ell = C_{\Omega/2,\ell}/ \sqrt{C_{\Omega, \Omega/2}}&  \label{4} \\
& |\ell \rangle_p \otimes |\ell \rangle_h = \frac{1}{{Z_{\ell, \Omega/2} }}
\; \left( A^\dagger B \right)^\ell |FS \rangle &  \nonumber
\end{eqnarray}

Performing the p-h transformation $B \rightarrow B^\dagger$, we deduce that (%
\ref{2}) can be written as a sum over the tensor product of particle and
holes states with amplitude $\psi_\ell$, where $\ell$ is the associated
occupation number. Tracing over the hole states one obtains from (\ref{4}) a
density matrix whose eigenstates are the particle states $|\ell\rangle_p$
with eigenvalues $w_\ell = \psi_\ell^2 \; (\ell = 0, 1, \dots, \Omega/2)$.
Tracing over the particle states yield identical results for the hole
states. In both cases the eigenvalues of the density matrix follow the
hypergeometric distribution $w_\ell = C_{\Omega/2,\ell}^2/ C_{\Omega,
\Omega/2}$ which for large values of $\Omega$ becomes a normal distribution
centered at $\Omega/4$ with quadratic deviation $\sigma= \sqrt{\Omega/2}$.
This is an interesting result for it implies that the PBCS state (\ref{2})
can be approximated to a great accuracy with a number of particle and hole
states of order $\sqrt{\Omega/2}$. We expect this result to hold for generic
PBCS states. % with amplitudes $v_j$ and $u_j$ ( $u_j^2 + v_j^2 =
%$1) $. 
%In this case the most probable particle or hole state would have an
%occupation number given by $\sum_{j=\Omega/2+1}^{\Omega} v_j^2$.

The gaussian decay of the weights $w_\ell$ of the density matrix offers an
ideal situation for the application of the DMRG. The DMRG works very well in
the cases where the weights decay exponentially ( see \cite{Okunishi} for
other types of decays). The p-h breaking allows for a smooth evolution of
the system from a few electron regime into a superconducting one.

Before we introduce the DMRG it is convenient to perform the following
canonical transformation,

\begin{equation}
c_{j,\sigma} = \left\{ 
\begin{array}{ll}
b^\dagger_{\Omega/2 +1 -j,\-\sigma} & j=1, \dots,\Omega/2 \\ 
a_{j-\Omega/2, \sigma} & j=\Omega/2+1, \Omega
\end{array}
\right.  \label{5}
\end{equation}

\noindent where the operators $a^\dagger_{j, \sigma}$ ( resp. $%
b^\dagger_{j,\sigma}$) create particles ( resp. holes) acting on the Fermi
sea (\ref{3}). Choosing the chemical potential $\mu$ as

\begin{equation}
\mu = \frac{d}{2} ( \Omega +1-\lambda)  \label{6}
\end{equation}

\noindent the Hamiltonian (\ref{1}) has the p-h symmetry $a_{j,\sigma}
\leftrightarrow b_{j, \sigma}$, and it can be written as,

\begin{eqnarray}
& H/d = K^A + K^B - \lambda ( A^\dagger A + B^\dagger B + A B + A^\dagger
B^\dagger) &  \label{7} \\
& K^A = \sum_{j=1, \sigma=\pm}^{\Omega/2} \tilde{\epsilon}_j
a^\dagger_{j,\sigma} a_{j,\sigma}, \;\; \; (\tilde{\epsilon}_j= j- \frac{1}{2%
} + \frac{\lambda}{2} ) &  \label{8} \\
& A^\dagger = \sum_{i=1}^{\Omega/2} a^\dagger_{i,+} a^\dagger_{i,-}&
\label{9}
\end{eqnarray}

\noindent where $K^B$ and $B$ can be obtained from $K^A$ and $A$ by the p-h
transformation $a_{i,\sigma} \leftrightarrow b_{i,\sigma}$. We have
substracted in (\ref{7}) the constant term $- \left( \frac{\Omega}{2}
\right)^2$, which is the energy of Fermi sea $|FS\rangle$, so that the
lowest eigenvalue of $H$ gives directly the condensation energy for even
grains $E^C_0 = \langle \psi | H | \psi \rangle - \langle FS | H | FS \rangle
$. For odd grains the level located at the Fermi sea is blocked and the
condensation energy can be computed following the same steps as in the even
case. From now on we shall concentrate on the latter case. We are now ready
to apply the DMRG to find the GS of the Hamiltonian (\ref{7}). At half
filling this state takes the generic form

\begin{equation}
|\psi \rangle = \sum_{\alpha, \beta} \psi_{\alpha,\beta}\; |\alpha\rangle_p
\otimes |\beta\rangle_h  \label{10}
\end{equation}

\noindent where the particle state $|\alpha \rangle _{p}$ must have a number
of particles $N_{p}$ equal to the number of holes $N_{h}$ of $|\beta
\rangle_h $ for the GS state (\ref{10}) to be non vanishing. The DMRG is an
algorithm that gives an optimal choice for the set of particle and hole
states entering in (\ref{10}). This set is constructed in successive steps
starting from small grains. We begin with a system with $\Omega =4$ energy
levels, which are chosen as the closest two particle and hole states near $%
\epsilon _{F}$. This system can be represented as $\bullet \bullet \circ
\circ $, where $\bullet $ stands for a particle level, while $\circ $ stands
for a hole one. The Fermi energy lies in between the $\bullet $'s and the $%
\circ $'s. The next step is to look for the GS of the Hamiltonian (\ref{7})
for $\Omega =4$ in the sector $N_{p}=N_{h}$. From the knowledge of $\psi
_{\alpha ,\beta }$ we define the reduce density matrix for the particle
subspace

\begin{equation}
\rho_{\alpha,\alpha^{\prime}}^{A}= \sum_{\beta} \psi^*_{\alpha,\beta}
\psi_{\alpha^{\prime}, \beta}  \label{11}
\end{equation}

The p-h symmetry implies that the corresponding density matrix in the hole
subspace coincides with (\ref{11}). The particle states which contribute the
most to the GS (\ref{10}) are the eigenvectors of the density matrix $%
\rho^{A}$ with highest eigenvalues $w_p$ \cite{DMRG}. For the system $%
\bullet \bullet \circ \circ$ we can work with all the eigenstates of $%
\rho^{A}$, but in general we shall only be able to keep the $m$ most
probable ones. With the information gained previously one builds the system
with $\Omega=6$. The general rule is to build the system with $\Omega= 2(n+1)
$ levels out of the system with $\Omega= 2n$. This is achieved by
constructing the system with $\Omega=2(n+1)$ as a superblock of the form $%
\bullet {\cal {A}}_n {\cal {B}}_n \circ$, where ${\cal {A}}_n$ (resp. ${\cal 
{B}}_n$) is the block which gives an effective description of the lowest $n$
particle ( resp. hole) levels in terms of $m$ states, while $\bullet$ and $%
\circ$ represent the $(n+1)^{{\rm th}}$ particle and hole levels added to
enlarge the system size.

The Hamiltonian $H_{\bullet {A} {B} \circ}$ of the superblock $\bullet {\cal 
{A}}_n {\cal {B}}_n \circ$ is given by

\begin{eqnarray}
& H_{\bullet {A} {B} \circ} = H_A + H_B + H_\bullet + H_\circ &  \nonumber \\
& + H_{AB} + H_{\bullet A} + H_{A \circ} + H_{\bullet B} + H_{B \circ} +
H_{\bullet \circ} &  \label{12}
\end{eqnarray}

\begin{equation}
\begin{array}{ll}
H_A & = K^{A}_n - \lambda A^\dagger_n A_n \\ 
H_\bullet & = \sum_\sigma \tilde{\epsilon}_{n+1} a^\dagger_{n+1,\sigma}
a_{n+1, \sigma} - \lambda a^\dagger_{n+1} a_{n+1} \\ 
H_{AB} & = -\lambda( A_n B_n + h.c.) \\ 
H_{ \bullet A} & = - \lambda( A_n a^\dagger_{n+1} + h.c.) \\ 
H_{A \circ} & = -\lambda( A_n b_{n+1} + h.c. ) \\ 
H_{\bullet \circ} & = -\lambda ( a_{n+1} b_{n+1} +h.c.)
\end{array}
\label{13}
\end{equation}

\noindent where $a_{i}^{\dagger }=a_{i,+}^{\dagger }a_{i,-}^{\dagger }$ and $%
A_{n},K_{n}^{A}$ are the operators $A$ and $K^{A}$ defined in eqs.(\ref{8})
and (\ref{9}) but with $\Omega /2$ replaced by $n$. $K^{(B)},b_{i}$ and $%
B_{n}$ have similar definitions. The terms $H_{B},H_{\circ },H_{\bullet B}$
and $H_{B\circ }$ can be derived from (\ref{13}) by the p-h transformation $%
A_{n}\leftrightarrow B_{n},a_{i}\leftrightarrow b_{i}$. The splitting (\ref
{12}) of the superblock Hamiltonian $H_{\bullet {A}{B}\circ }$ recalls the
one used in the momentum space DMRG \cite{Xiang}. However the latter
reference uses a finite system algorithm which does not exploits the p-h
symmetry. The DMRG provides a many body description of the blocks ${\cal A}%
_{n}$ and ${\cal B}_{n}$, which means that the operators acting on these
blocks are represented by $m\times m$ matrices. In our case the operators
that we need to keep track are $[A_{n}]$, $[A_{n}^{\dagger }A_{n}]$ and $%
[a_{j,\sigma }^{\dagger }a_{j,\sigma }]$. Given these operators we can
construct the superblock Hamiltonian (\ref{13}) and look for the GS in the
sector $N_{p}=N_{h}$ using the Lanczos method. The dimension of this Hilbert
space ( ${\rm dim}{\cal H}_{\Omega ,m}$) is smaller than $4m^{2}$, for the
constraint $N_{p}=N_{h}$ eliminates the states away from half filling. ${\rm %
dim}{\cal H}_{\Omega ,m}$ is usually much smaller than the exact dimension
of the Hilbert space of states with $\Omega $ levels at half filling which
is given by the combinatorial number $C_{\Omega ,\Omega /2}$.

Given the GS of the superblock we obtain, using eq.(\ref{11}), the density
matrix of the particle system $\bullet {\cal A}_n$ and diagonalize it
keeping the $m$ most probable states with weight $w_p$. The error of the
truncation is measured by $1-P_m \; (P_m = \sum_{p=1}^m w_p)$. The latter
states form a new basis of $\bullet {\cal A}_n$ denoted as ${\cal A}_{n+1}$
and they give an effective description of the particle subspace with $n+1$
levels. The hole block ${\cal B}_{n+1}$  is a mirror image of the particle
block ${\cal A}_{n+1}$. The DMRG proposed above is an infinite system
algorithm, which is sufficient to study moderate system sizes ($N \le 400)$.
A way to improve the numerical accuracy of the infinite system method is to
choose an effective value of the coupling constant $\lambda_n({\rm bulk})$
at the $n^{{\rm th}}$ DMRG step in such a way that the value of the bulk gap
is the one of the final system with coupling constant $\lambda= \lambda({\rm %
bare}$). This is guaranteed by the equation

\begin{equation}
{\rm sinh}\frac{1}{\lambda _{n}({\rm bulk})}= \frac{2(n+1)}{\Omega }{\rm sinh%
}\frac{1}{\lambda({\rm bare}) }  \label{14}
\end{equation}

\noindent where $\Omega$ is the final number of levels to reach and $2(n+1)$
is the number of levels at each step. Let us now present our results. A DMRG
calculation for $\Omega = 24$ and $m=60$ agrees with the exact Lanczos
condensation energy obtained in \cite{MFF} in the first 9 digits. The
largest DMRG superblock matrix involved in the calculation is $3066$ to be
compared with the Lanczos matrix of dimension $2704156$. For $\Omega \leq 400
$ and $m=60$ the condensation energy is computed with a relative error less
than $10^{-4}$.

The numerical improvement achieved by the use of the effective coupling
constant $\lambda_n({\rm bulk})$ defined in (\ref{14}) as compared with the
use of $\lambda= \lambda({\rm bare})$ in the DMRG steps is illustrated in
table 1.

\begin{center}
\begin{tabular}{|c|c|c|c|c|}
\hline
$m$ & $E_0^{C}({\rm bare})/d$ & $E_0^C({\rm bulk})/d$ & $1-P_{m}$ & ${\rm dim%
}{\cal H}_{100,m}$ \\ \hline
50 & -40.44623 & -40.50014 & $2.0\times 10^{-9}$ & 2108 \\ 
70 & -40.48878 & -40.50068 & $7.1\times 10^{-11}$ & 3622 \\ 
90 & -40.49815 & -40.50074 & $1.1\times 10^{-11}$ & 6306 \\ 
110 & -40.49983 & -40.50075 & $1.5\times 10^{-12}$ & 9720 \\ \hline
\end{tabular}

Table 1. GS condensation energy $E^C_0$ for $\Omega =100$ and $\lambda = 0.4$
computed using the effective coupling constant and the bare one. $1-P_{m}$
is the truncation error of the last iteration and ${\rm dim} {\cal H}_{100,m}
$ is the largest dimension of the superblock.
\end{center}

Let us next consider the crossover between the weak and the strong coupling
regimes. The PBCS results of ref. \cite{Braun2} suggest a sharp crossover
between these two different regimes at characteristic level spacings $%
d_{0}^{C}=0.5\Delta $ and $d_{1}^{C}=0.25\Delta $. For $d<d_{b}^{C}$ the
condensation energy is an extensive quantity ($\sim 1/d$) corresponding to a
BCS-like behavior, while for $d>d_{b}^{C}$ this energy is an intensive
quantity (almost independent of $d$) \cite{Braun2}. In figure 1 we plot the
DMRG results together with those of reference \cite{Braun2}. From this
comparison it is apparent that the DMRG gives significant lower energies
than the PBCS ansatz. The DMRG is a variational method and in the region
under study we expect our results to coincide with the exact ones with a
relative error less than $10^{-4}$. Fig. 1 also shows that the crossover
between the BCS and the fluctuation dominated (f.d.) regimes takes place in
a region which is wider than the one predicted by the PBCS approach and that
there is no signs of critical level spacings $d_{b}^{C}$. A more
quantitative characterization of this crossover is obtained by fitting the
DMRG results to the formula ( see Fig. 1)

\begin{eqnarray}
&E_{b}^{C}/{\Delta }=-c_{1}/{\rm ln}(1+c_{2}\frac{d}{\Delta })+c_{3}+c_{4}%
\frac{d}{\Delta }&  \label{15} \\
&c_{1}=1.48,c_{2}=3.05,c_{3}=-1.98,c_{4}=0.08 ,\; (b=0)&  \nonumber \\
&c_1 = 0.36, c_2 =0.86, c_3=-1.95,c_4=0.16,\; (b=1) &  \nonumber
\end{eqnarray}

%%%%%%%%%%%%%%%%%%%%%%%%%%%%%%%%%%%%%%%%%%%%%%%%
\begin{figure}
\vspace{2.0cm}
\hspace{-0.1cm}
\epsfxsize=7cm \epsffile{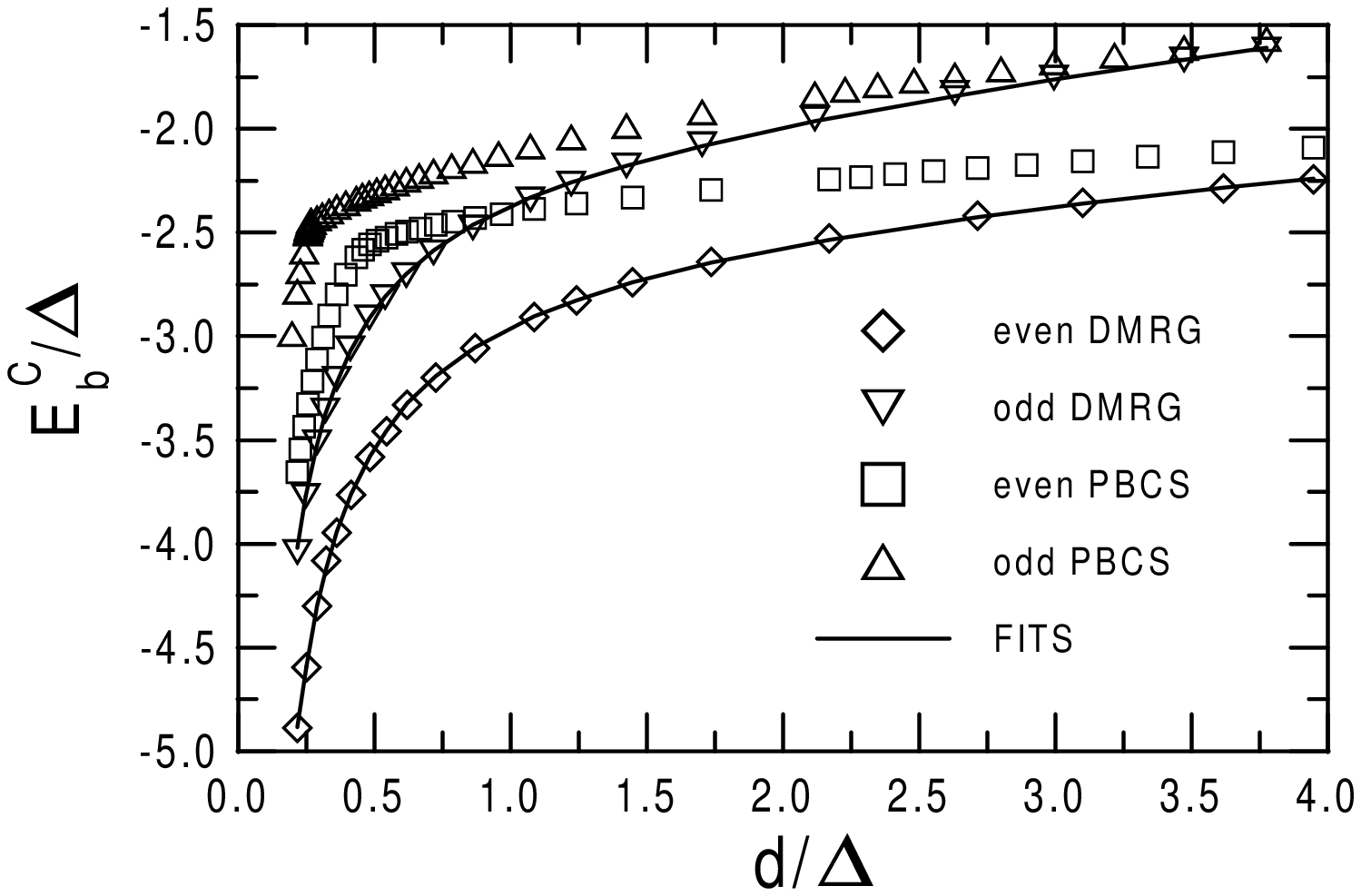}
%\vspace{-3.0cm}
\narrowtext
\caption[]{GS condensation energies $E^C_b (b=0,1)$ as a function of $d/\Delta$
for $\lambda=0.224.$. $\Omega$ ranges from 22 ( resp. 23) up to 400 ( resp.
401) for even ( resp. odd) grains and $m=60$. The PBCS results are those of
ref. \protect\cite{Braun2}. }
\label{fig1}
\end{figure}
%%%%%%%%%%%%%%%%%%%%%%%%%%%%%%%%%%%%%%%%%%%%%%%%%

\noindent which interpolates between the bulk-BCS like behavior, given by $%
E^{C}_0=-\frac{c_{1}}{c_{2}}\Delta ^{2}/d$ ( $d/\Delta <<1$), and the f.d.
regime ($d/\Delta >>1)$ characterized by logarithmic corrections \cite{ML}.
Indeed from (\ref{15}) we get $c_{2}/c_{1}=2.06$ for $b=0$ and $c_2/c_1= 2.4$
for $b=1$ which are both close to the bulk value given by $2$ \cite{BvD}.

The previous results are consistent with the probabilities of the $m$ states
kept by the DMRG as a function of the number of particles (resp. holes) of
the states $|\alpha \rangle _{p}$ ( resp. $|\beta \rangle _{h}$) appearing
in (\ref{10}), which we denote as $N_{p-h}$. Since the DMRG keeps several
states with the same value of particles and holes, it is appropriate to sum
the probabilities of all the states with the same value of $N_{p-h}$, which
we shall denote as $w(N_{p-h})$. Each value of $w(N_{p-h})$ is dominated by
a single state which contributes the most. In Fig.2 we plot $w(N_{p-h})$ in
the case $\lambda =0.224$ and several values of $\Omega $, which correspond
to those plotted in Fig.1. The rapid decay of the weights recalls the
gaussian decay of the eigenvalues of the density matrix of the PBCS (\ref{2}%
).

For $\Omega=22,100,180,270$ and $400$ the most probable states have $%
N_{p-h}=0,0,1,1,2$ respectively, while the next two most probable states
have occupation numbers $|N_{p-h}\pm 1|$. As $\Omega$ increases the most
probable state moves to higher values of $N_{p-h}$, becoming eventually
commensurable with $\Omega$ in the extreme BCS regime. It may seem from
these results that the p-h DMRG is not capable to describe the bulk-BCS
regime. This is not so for we have indeed observed this regime for higher
values of $\lambda$. 
%Recall also the discussion at the beginning of this
%letter where we showed analytically that the PBCS state (\ref{2}) can be
%properly described by a set of particle and hole states with dimension of
%order $\sqrt{\Omega /2}$.

In summary, we have proposed in this letter a new version of the DMRG which
is suitable to study the GS properties of the BCS pairing Hamiltonian of the
ultrasmall superconducting grains. 
We believe that the p-h DMRG formulated in
this work can be applied to a more general variety 
of fermionic  systems in Condensed
Matter, Atomic, Molecular \cite{mole} and Nuclear Physics.
Performing  a Hartree-Fock (HF)
transformation we can look for  the best single particle
basis to begin  the DMRG procedure.

%%%%%%%%%%%%%%%%%%%%%%%%%%%%%%%%%%%%%%%%%%%%%%%%
\begin{figure}
\vspace{2.0cm}
\hspace{-0.1cm}
\epsfxsize=6cm \epsffile{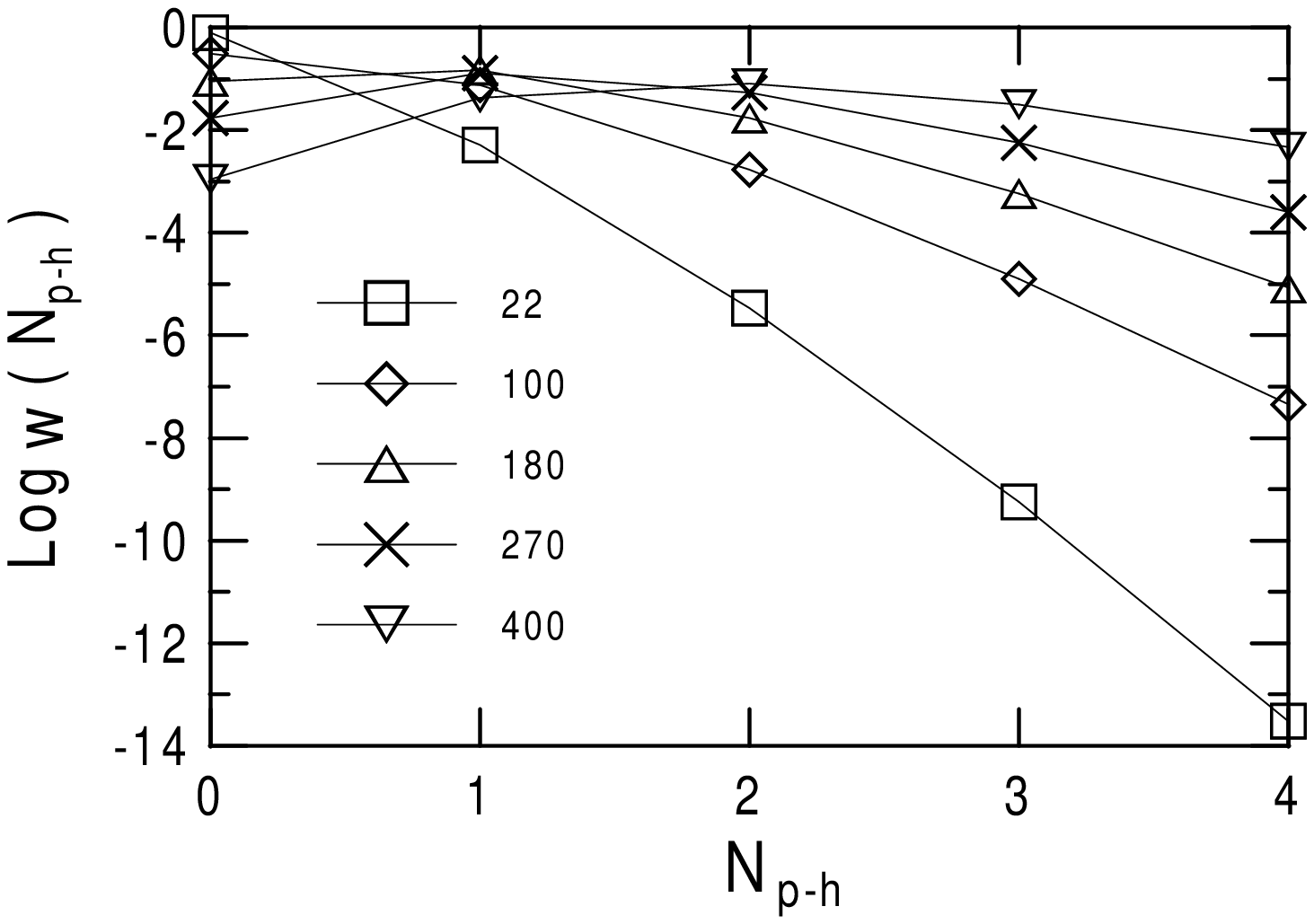}
%\vspace{-3.0cm}
\narrowtext
\caption[]{DMRG weights for $m=60$ and $\lambda = 0.224$ as a function of the
number of particle states for different values of $\Omega $.}
\label{fig2}
\end{figure}
%%%%%%%%%%%%%%%%%%%%%%%%%%%%%%%%%%%%%%%%%%%%%%%%%

The main limitation concerns  again the amount of important
states which in general will grow with the size of the system
to some power. For example 
in the pairing problem it grows with the square root
of the size which allows us to study large systems.

The comparison of the DMRG and the PBCS results \cite{Braun2} shows no signs
of critical level spacings separating qualitative different regimes. We
rather observe a smooth logarithmic-like crossover which contradicts the
sharp crossover predicted by the PBCS ansatz. This latter feature is an
artifact of the PBCS method which is unable to capture the true nature of
the crossover. In summary the fluctuations seem to play a major role for no
too small grains.

{\bf Acknowledgments} We thank M.A. Martin-Delgado, S. Pittel, P. Schuck and
A. Zuker for conversations and F. Braun and J. von Delft, for sending us
their numerical data. This work was supported by the DGES spanish grants
PB95-01123 (J.D.) and PB97-1190 (G.S.).

%%%%%%%%%%%%%%%%%%%%%%%%%%%%%%%%%%%%%%%

%%%%%%%%%%%%%%%%%%%%%%%%%%%%%%%%%%%%%%%

\end{document}